# Super-ideal diodes at the Schottky-Mott limit in gated graphene-WSe$_2$ heterojunctions

S. W. LaGasse[1], P. Dhakras[1], T. Taniguchi[2], K. Watanabe[2], J. U. Lee[1]*

[1] Colleges of Nanoscale Science and Engineering, State University of New York Polytechnic Institute, Albany, New York 12203, USA

[2] National Institute for Materials Science, 1-1 Namiki, Tsukuba 305-0044, Japan.

*Correspondence to: JLee1@sunypoly.edu

**Abstract:** Metal-semiconductor interfaces, known as Schottky junctions, have long been hindered by defects and impurities. Such imperfections dominate the electrical characteristics of the junction by pinning the metal Fermi energy. We report measurements on a boron nitride encapsulated graphene-tungsten diselenide (WSe$_2$) Schottky junction which exhibits ideal diode characteristics and a complete lack of Fermi-level pinning. The Schottky barrier height of the device is rigidly tuned by electrostatic gating of the WSe$_2$, enabling experimental verification of the Schottky-Mott limit in a single device. Utilizing this exceptional gate control, we realize a "super-ideal" gated-Schottky diode which surpasses the ideal diode limit. Our results provide a pathway for defect-free electrical contact to two-dimensional semiconductors and open up possibilities for circuits with efficient switching characteristics and higher efficiency optoelectronic devices.

Schottky junctions (SJs), which are formed at a metal-semiconductor (M-S) interface, are characterized by a current rectifying energy barrier. Ideally, the barrier is determined by only the metal work function and semiconductor electron affinity, in a case known as the Schottky-Mott (SM) limit. Typically, however, defect states at the M-S interface induce Fermi-level pinning (FLP) in the metal and dictate the energy barrier height (*1*). Experiment has approached the SM limit in two dimensional (2D) semiconductors contacted with three-dimensional metal contacts (*2*, *3*). However, despite theoretical predictions (*4*, *5*), experimental observation of the SM limit using 2D metals has been illusive (*6*–*19*).

Van der Waals (vdW) heterostructures (*20*), especially when passivated with hexagonal boron nitride (*h*-BN) (*21*, *22*), present an excellent platform for studying the SM limit. Graphene (*20*, 21), a semimetal with a gate-tunable work function (23), is a promising alternative to traditional bulk metal electrical contacts to 2D semiconductors (15). In lieu of using different metals, we propose a modified Schottky-Mott rule for gate-tunable SJs in which the gate voltage ($V_G$) directly modulates the barrier height ($\Phi_B$), $\left|\frac{d\Phi_B}{dV_G}\right| = S_\text{G}$. When $S_\text{G} = 1$, the system is operating at the Schottky-Mott limit. Here, we present measurements on a gated graphene-WSe$_2$ SJ for which $S_\text{G} \approx 1$.

Electrical transport across SBs may be modeled by Shockley's diode equation, $I_D = I_0\left[e^{qV_D/nk_BT} - 1\right]$, where $I_\text{D}$, $I_0$, $q$, $V_\text{D}$, $n$, $k_\text{B}$, and $T$ are the drain current, reverse bias leakage current, electron charge, applied drain voltage, ideality factor, Boltzmann's constant, and temperature, respectively. W. Shockley himself described his diode equation as a formula giving the maximum rectification of charge (22). Indeed, the ideal diode equation, which is characterized by an ideality factor $n = 1$, is one of the most non-linear equations to describe a physical system. Due to the cleanliness of the graphene-WSe2 interface studied here, our junction exhibits nearly ideal diode characteristics. By exploiting our one-to-one electrical control over the Schottky barrier (SB), we also demonstrate a tunable effective ideality factor which exceeds the ideal diode limit.

An extremely clean graphene-WSe2 interface is enabled by a high-temperature version of standard dry-stamping techniques in which our device channel is passivated by encapsulation in h-BN (22, 27, 28). The device is selectively etched to create a graphene-WSe2-graphene channel, and electrical contact to the graphene is made by one-dimensional edge contacts (26). In order to study the electrical transport properties of just one graphene-WSe2 interface, our device is positioned over two buried split-gates (Fig. 1A). A three-dimensional schematic of our device is depicted in Fig. 1B. One split-gate, $V_{\text{G2}}$, may be tuned such that the graphene-WSe2 interface on one side of the device is electrically transparent. Meanwhile, the other split-gate ($V_{\text{G1}}$) is used to tune the SB height at the other graphene-WSe2 interface (Fig. 1C).

Fig. 2A shows the $I_\text{D}$-$V_\text{D}$ characteristic of our device for a varying $V_{\text{G1}}$ between 0V and 1V (lightly *p*-type) with a fixed $V_{\text{G2}}$ of -10V (heavily *p*-type). For this configuration we observe a nearly-ideal *p*-type Schottky barrier. By applying an increasingly positive gate voltage on $V_{G1}$, the measured forward bias current uniformly shifts towards larger values of $V_D$, while the reverse bias leakage current rapidly drops below measurable levels.

Both the reverse and forward bias characteristics of our device show striking agreement with the ideal diode equation. In particular, the reverse bias leakage current is nearly constant with the applied drain voltage in a way which is rarely observed. Due to the ideal behavior displayed in our measurement, we can fit the forward bias current of the device at each $V_{G1}$ in order to extract the device ideality factor and extrapolate the reverse bias leakage current (Fig. S1). In Fig. 2B the resultant fitting indicates near-ideal diode behavior with *n*~1.1-1.2 and exponentially decaying reverse bias leakage current, tuned across a 1V range of gate voltages.

From Fig. 2b, we extract an approximate thirteen orders of magnitude change in the leakage current in response to varying $V_{G1}$, nearing the 60 mV/decade limit. Indeed, the globally-gated transfer curve of our device also indicates near-60 mV/decade characteristics (Fig. S2). Motivated by this result, we performed a temperature-dependent study (Fig. S3) to determine the SB height as a function of $V_{G1}$ (Fig. 3). Our results indicate that there is a fundamental lack of FLP present at the interface between graphene and WSe$_2$ in our device, displaying gate-tuned, Schottky-Mott limited, characteristics with $S_G \approx 1$.

Our exceptional electronic gate control over the Gr-WSe2 Schottky barrier height is analogous to an ideal metal-oxide-semiconductor field effect transistor (MOSFET) operated in the sub-threshold regime. In a similar manner to how the gate voltage directly couples to the electric potential of the channel in a MOSFET, our split-gate voltage directly modulates the height of the graphene-WSe2 Schottky barrier (S2.1). This same behavior yields the extremely flat reverse bias leakage current observed in the device.

We derive (S2.2) a modified, gate voltage-dependent ideal diode equation, for Schottky-Mott limited junctions, $I_D = A^* A T^{3/2} e^{-q(\Phi_0 + V_{G1})/k_B T} \left[ e^{qV_D/nk_B T} - 1 \right]$, where $A^*$ is the effective Richardson constant, $A$ is the effective SJ area, and $\Phi_0$ is the SB height at $V_{G1} = 0$ V. The gate voltage directly controls the SB height at the graphene-WSe2 interface and the drain voltage forward or reverse biases the diode. Atypically, the gate and drain voltages act at parity to determine the $I_D$-$V_D$ characteristics of the device.

Normally, the $I_D$-$V_D$ characteristics of a diode are strongly determined by $n$. Here, the additional control provided by the gate allows some startling results. By sweeping the gate and drain voltages simultaneously, we impose digital control over the ideality constant of the Schottky diode. The drain current in such a measurement is described by the ideal diode equation, but the ideality factor is replaced with an effective ideality factor (S2.2):

$$n \to n_{\text{Eff}} = \frac{n}{1 - nm} \quad (1)$$

Here, $n_{\text{Eff}}$ is the new ideality factor, $m$ is the rate of change of $V_{G1}$ relative to $V_D$, and $n$ is the orginal diode ideality factor. Going against convention, in our three-terminal measurement, $n_{\text{Eff}}$ can be tuned to be less than one, in what we are calling a "super-ideal" configuration.

Fig. 4 shows diode measurements made for a starting split-gate voltage $V_{G1} = 0.4V$. By sweeping $V_{G1}$ in the opposite direction of $V_D$ ($m < 0$), we achieve super-ideal diode characteristics down to $n_{\text{Eff}} = n/10$. By sweeping the gate voltage in the same direction as the drain voltage, we observe up to $n_{\text{Eff}} = 4n$. Similar measurements are shown for different starting $V_{G1}$ voltages (Fig. S5). Physically, by simultaneously shrinking the height of the graphene-WSe$_2$ SB (by tuning $V_{G1}$) and sweeping the drain, we observe super-ideal diode characteristics. Conversely, increasing the height of the graphene-WSe$_2$ SB during the drain voltage sweep yields $n_{\text{Eff}} > n$ (Fig. S6). Our measurements show strong agreement with Eq. 1 (Fig. S7). In fact, the only limitation we encountered in how small we can make $n_{\text{Eff}}$ was in the resolution of the voltage sources we employed.

Our results provide verification of the SM rule in a single device. We provide an avenue for studying the SM limit in gated SJs, circumventing the requirement for fabricating many separate devices. The ability to create unpinned graphene-2D semiconductor junctions within the existing vdW heterostructure framework will enable researchers to probe exotic physics requiring high-quality electrical contact. Furthermore, our gated-Schottky diodes result in a tunable effective ideality factor. Tuning $n_{\text{Eff}} < 1$ will enable new circuits with efficient switching characteristics. Finally, tuning $n_{\text{Eff}} > 1$ whilst minimizing the reverse bias leakage current is promising for creating higher efficiency PV devices.

**Acknowledgments:** We thank L. Wang, P. Sutter, Y. Huang, and A. Lombardo for useful discussions, M.T. Murphy for assistance with formulation of the polycarbonate stamps used in device fabrication, and B. Taylor, S. Stewart, S. Warfield, and K. Unser for technical support.

**Funding:** This work is supported by the U.S. Naval Research Laboratory through grant number N00173141G017 and National Science Foundation through grant number 1606659.

**Supplementary Materials:**

Materials and Methods

Supplementary Text

Figures S1-S9

References (*29-34*)

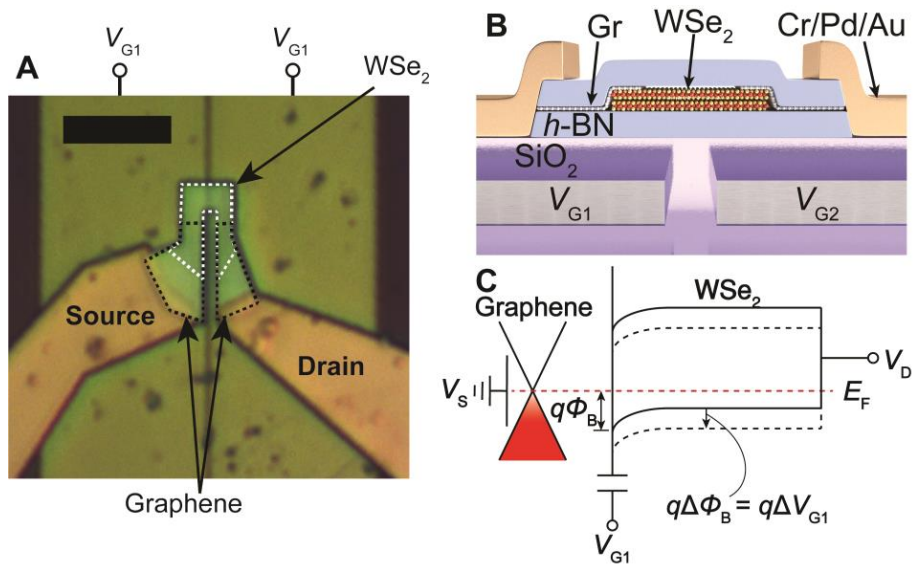

**Fig. 1.** Gated graphene-WSe$_2$ SJ. (**A**) Optical micrograph of a gated graphene-WSe$_2$ SJ encapsulated in *h*-BN. The polysilicon split-gates have a 100 nm spacing are buried under 100 nm of SiO$_2$. The scale bar is 10 µm. (**B**) Schematic of a gated SJ. Buried polysilicon split-gates are used to configure the device such that one Gr-WSe$_2$ junction is electrically transparent (using $V_{G2}$) and the other is a gate-tunable Schottky diode (using $V_{G1}$). (**C**) Energy band diagram depicting tuning of the SJ. Gate one is capacitively coupled to the channel and modulates the Schottky barrier. When the semiconductor is lightly doped, a change $\Delta V$ in the gate voltage exactly modulates the Schottky barrier height by $q\Delta V$.

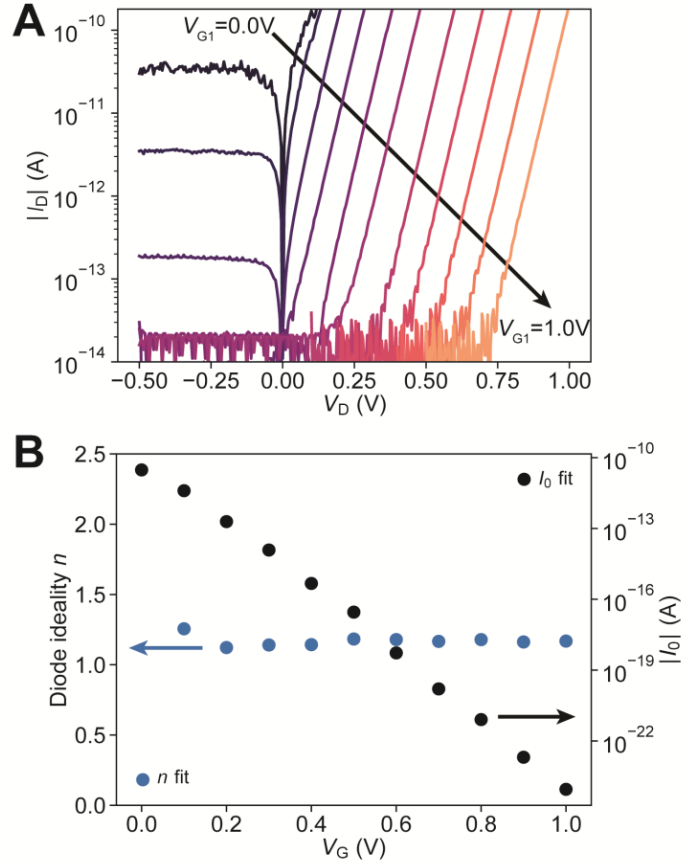

**Fig. 2.** (**A**) $I_D$-$V_D$ characteristic of the gated SJ. $V_{G2}$ is fixed to -10V and $V_D$ sweeps are made for $V_{G1}$ set between 0.0V and 1.0V. For values of $V_{G1}$>0.2V, only a portion of the forward bias current is large enough to be measured above our noise limit of $2\times10^{-14}$A. (**B**) Diode ideality constant $n$ and reverse bias leakage current $I_0$ as a function $V_{G1}$ of obtained by fitting the results in Fig. 2A to the diode equation.

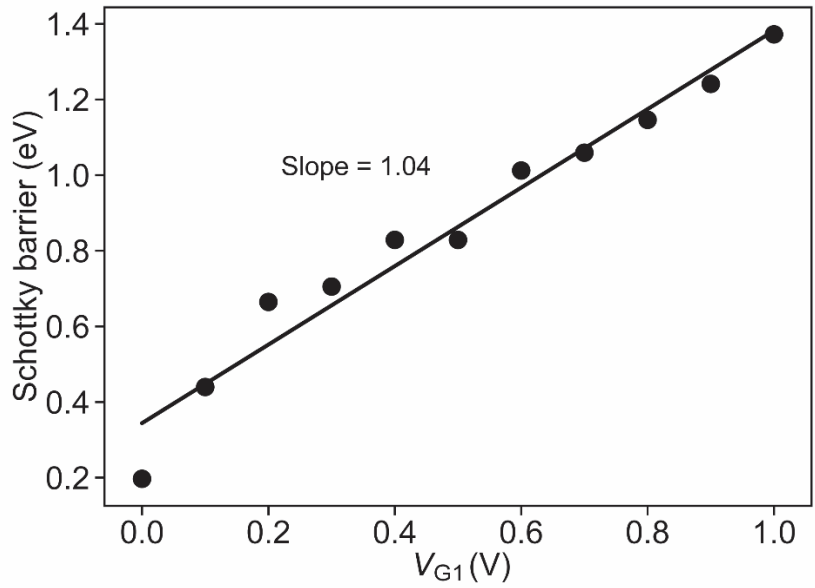

**Fig. 3.** Probing the Schottky-Mott limit in a gated Gr-WSe$_2$ SJ. The Schottky barrier height is determined by fitting to the 2D Schottky diode thermionic emission relationship. $S_G \approx 1$, indicating gate one has Schottky-Mott limited control over the Schottky barrier height.

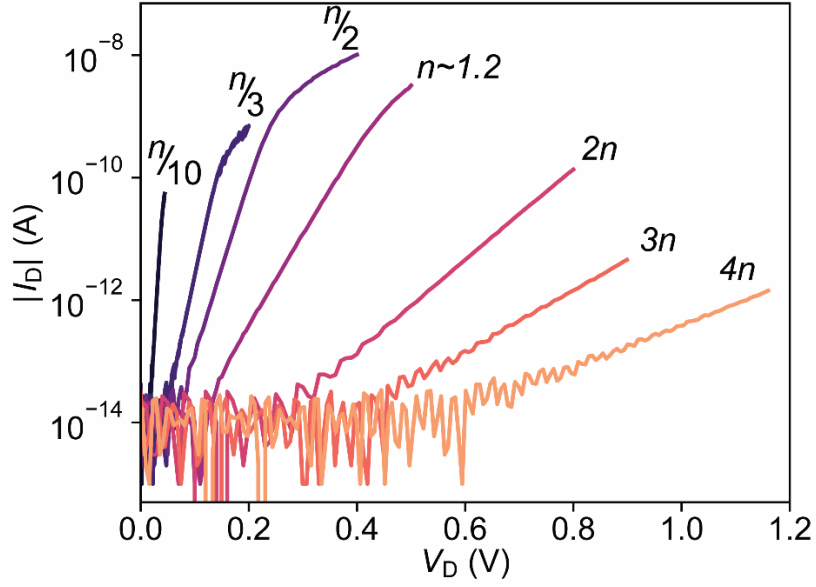

**Fig. 4.** Super-ideal diode measurements. Simultaneous sweeps of $V_{G1}$ and $V_D$ are performed, where $V_{G1} = V_0 + mV_D$. The starting value of $V_{G1}$, $V_0$, is 0.4V. Tuning the value of $m$ yields effective ideality constants between $n/10$ and $4n$, where $n \approx 1.2$.

# Supplementary Materials for

## Super-ideal diodes at the Schottky-Mott limit in graphene-WSe$_2$ heterojunctions

S. W. LaGasse, P. Dhakras, T. Taniguchi, K. Watanabe, J. U. Lee

Correspondence to: JLee1@sunypoly.edu

**This PDF file includes:**

    Materials and Methods
    Supplementary Text
    Figs. S1 to S9

**Materials and Methods**

1.1 <u>Modified van der Waals heterostructure assembly</u>

The $h$-BN encapsulated Gr-WSe$_2$ vdW heterostructure studied in this work is assembled using a modified dry transfer technique which takes influence from several published results in the literature (*22, 27, 28*). The basis of the dry transfer technique is as follows: a polymer stamp is used to pick up a top layer of $h$-BN. The top $h$-BN is then used in subsequent pick-ups to form a bottom $h$-BN/WSe$_2$/Gr/top $h$-BN heterostructure in which the active layers (WSe$_2$/Gr) are never exposed to polymer residue.

To fabricate the stamp used in our vdW heterostructure we began by casting SYLGARD 184 polydimethylsiloxane (PDMS) in a ratio of 9:1 parts base to curing agent onto acetone/isopropyl alcohol (IPA) washed silicon wafers, creating a roughly 1 mm layer. A razor blade is used to cut out a few millimeter square block of the PDMS which is mounted to a glass slide. Separately, polycarbonate (PC, Goodfellow) is dispersed 10% by weight in chloroform and is cast as in Ref. (*27*) by putting drops of PC between two glass slides and sliding them apart. Once dried, the PC film is peeled off the glass slide by an adhesive tape window forming a free-standing PC membrane. The membrane is then applied over the PDMS block to form the basis of our stamp.

As created, the PC stamp is very sticky near 40 °C, enabling pick-up of the different flakes used in the vdW heterostructure. Generally, we find increased temperature results in a more sticky PC stamp, up until the PC delaminates from the PDMS block or the PC melts. While the PC is very sticky, the contact of the stamp proceeds in a very jumpy manner, which is detrimental to forming clean heterostructures. This characteristic forces heterostructure with standard PC stamps to be performed at temperatures less than 40 °C. However, we found that by doing a three-minute hot-plate anneal of the finished stamp at 200 °C (about 50 °C higher than the glass transition temperature of PC), the PC reflows around the PDMS block, significantly altering the properties of the stamp. Our heat treated stamps display very smooth action while layering the heterostructure up to about 120 °C and become sticky at around 125 °C, enabling pick-up of crystals. By assembling our heterostructures at elevated temperatures, fewer interfacial bubbles form, as others have noted (*29*). Finally, once a vdW heterostructure has been assembled on a stamp, it may be placed in a desired location, and the PC is melted to the sample at 230-250 °C.

1.2 Materials

The graphene, few-layer WSe$_2$, and *h*-BN used in this work are isolated from bulk crystals by mechanical exfoliation using standard adhesive tape. To obtain large area exfoliated crystals, we have found the O$_2$ plasma cleaning technique of Huang *et al.* to be useful (*30*). Graphene crystals are exfoliated from commercial bulk graphite (NGS Naturagraphit "graphenium"). Few-layer WSe$_2$ crystals are exfoliated from commercial WSe$_2$ (2D Semiconductors). The bulk *h-BN* crystals were grown using a high temperature and pressure technique (*31*).

1.3 Device fabrication

We begin by assembling an *h*-BN/WSe$_2$/Gr/top *h*-BN heterostructure as described above. Using optical contrast, we select ~20 nm thick *h*-BN, few-layer WSe$_2$, and monolayer graphene. During assembly, graphene is picked up such that it only covers a portion of the top *h*-BN. Importantly, in the subsequent pickup WSe$_2$ is partially covered by the graphene. The heterostructure is positioned over a substrate consisting of a pair of split-gates with 100 nm separation. The split-gates have a 100 nm spacing are buried underneath 100 nm of deposited gate SiO2, the fabrication of which was performed in the 300 mm chip fabrication line at the College of Nanoscale Science and Engineering in Albany, NY and has been described elsewhere (*32*). Using an optical microscope, we align the heterostructure so that the gap between the split-gates runs down the middle of the $WSe_2$ flake. The stamp is laminated to the final substrate, and the PC is removed in boiling chloroform, leaving behind the heterostructure over the split-gates.

We use standard electron-beam lithography and etching techniques to define the geometry vdW heterostructure after transfer. Polymethyl methacrylate 950K dispersed 4% by weight in anisole (Microchem) electron-beam resist is spun onto the sample to achieve about a 200 nm thick layer. We expose the sample using a Raith Voyager system with a 50 kV acceleration voltage. Development of the PMMA mask is subsequently performed in a 1:3 methyl isobutyl ketone:IPA solution. Reactive ion etching (RIE) in an $SF_6/O_2$ plasma is performed using a Technics 800 Micro RIE to etch through the vdW heterostructure, defining the channel. By careful patterning, we note that it is possible to define multiple contacts to the $WSe_2$ channel using a single piece of graphene, as was previously demonstrated for $MoS_2$-based devices (*33*). After the device shape has been defined, a subsequent patterning step is performed in the same way as before to make electrical edge contacts to the graphene (*22*). Prior to evaporating metal contacts, a short RIE step (using the same recipe, but for less time) is performed, to remove any polymer residue which may inhibit the formation of edge contacts. A 3/15/60 nm Cr/Pd/Au stack is next evaporated at rates exceeding 1 Å/s. Finally, metal lift-off is performed in boiling acetone and the finished device is rinsed in acetone/IPA prior to being measured.

1.4 Measurement methods

Measurements are performed in vacuum in a Lakeshore CPX-VF cryogenic probe station. The probe station is equipped with a hot stage which we used to extract the Schottky barrier height. Electrical measurements are performed using an HP 4156B semiconductor parameter analyzer.

1.5 Determination of Schottky barrier height

The Schottky barrier heights shown in Fig. 3 of the main text are extracted from $I_D$-$V_D$ measurements using the 2D thermionic emission model for Schottky barriers:

$$I_D = AA^*T^{3/2}e^{-q\Phi_B/k_BT}\left[e^{qV_D/nk_BT} - 1\right] \tag{S1}$$

$A$, $A^*$, $T$, $q$, $\Phi_B$, $k_B$, $V_D$, and $n$ are the the effective junction area, the Richardson constant, temperature, electron charge, Schottky barrier height, Boltzmann's constant, drain voltage, and diode ideality factor, respectively. By reverse biasing the Schottky junction, in this case by applying a negative drain bias, Eq. S1 reduces to $I_D = I_0 = AA^*T^{3/2}e^{-q\Phi_B/k_BT}$. From here, a simple Arrhenius relation may be obtained to directly extract $\Phi_B$:

$$\ln\left(\frac{I_0}{T^{3/2}}\right) = \ln(AA^*) - \frac{q\Phi_B}{k_BT} \tag{S2}$$

In order to determine the reverse bias leakage current in Eq. S2 for the configurations in which $I_0$ is too small to be directly measured, we fit the forward bias portion of the measurement to the diode equation. Such a fit has previously been performed to extract extremely low leakage currents from near-ideal carbon nanotube diodes (*34*). Results from our fitting procedure for selected data from Fig. 2a of the main text are presented in Fig. S1.

Fig. S2a shows temperature-dependent measurements performed on the graphene-WSe$_2$ GSD discussed in the main text. Using fitted $I_0$ values for each split-gate configuration, we obtain Fig. S2b from Eq. S2. By performing a linear fit of $\ln\left(\frac{I_0}{T^{3/2}}\right)$ a function of $\frac{1}{k_\text{B}T}$, the Schottky barrier height for each value of $V_{G1}$ can be directly obtained from the slope of each curve.

**Supplementary Text**

<u>Capacitance model</u>

A capacitor-based voltage divider model analogous to a planar MOSFET operated in subthreshold (lightly doped) regime at the 60 mV/decade limit is employed to understand the Gr/WSe$_2$ GSD. The schematic for the model is depicted in Fig. S3, showing just the Gr/WSe$_2$ interface where the Schottky barrier is present. $V_{G1}$ is capacitively coupled to the channel through the capacitance $C_\text{G}$. Analogous to the bulk capacitance in a planar MOSFET, we include a capacitance between the device channel and outside environment, which we call $C_\text{Parasitic}$. Additionally, we assume the drain voltage $V_\text{D}$ rigidly couples to potential of the channel. For non-zero values of $V_\text{D}$, $V_{G1}$ relative to the channel potential must be replaced by an effective gate voltage $V_{\text{G, eff}} = V_{G1} - V_\text{D}$. Treating this system as a voltage divider leads to the following relationship for the channel potential, $\phi_\text{S}$:

$$\phi_\text{S} = \frac{C_\text{G}}{C_\text{G} + C_\text{Parasitic}} V_{\text{G, eff}} + V_\text{D} \tag{S3}$$

Since $C_\text{G} \gg C_\text{World}$, Eq. S3 may be reduced to $\phi_\text{S} = V_{\text{G, eff}} + V_\text{D}$, which may be reduced further to finally obtain

$$\phi_S = V_{G1}. \tag{S4}$$

Derivation of super-ideal diode equation

Starting from the standard thermionic emission model for a 2D Schottky barrier (*10*),

$$I_D = A^*AT^{3/2}e^{-q\Phi_B/k_BT}\left[e^{qV_D/nk_BT} - 1\right], \tag{S5}$$

and using Eq. S4, we rewrite the Schottky barrier height $\Phi_B$ as $\Phi_B = \Phi_0 + V_{G1}$, where $\Phi_0$ is the Schottky barrier height when $V_{G1} = 0V$. This substitution yields,

$$I_D = A^*AT^{3/2}e^{-q(\Phi_0+V_{G1})/k_BT}\left[e^{qV_D/nk_BT} - 1\right]. \tag{S6}$$

We modify Eq. S6 to account for the simultaneous gate voltage/drain voltage sweeps discussed in the main text by re-writing $V_{G1} = V_0 + mV_D$, where $V_0$ is the value of $V_{G1}$ when the measurement is started and $m$ is the rate at which the gate voltage is swept relative to the drain voltage. Making the substitution for $V_{G1}$ and rearranging the equations yields:

$$I_D = A^*AT^{3/2}e^{-q(\Phi_0+V_0)/k_BT}\left[\exp\left(\frac{q}{nk_BT}V_D - \frac{q}{k_BT}mV_D\right) - \exp\left(\frac{-q}{k_BT}mV_D\right)\right]. \tag{S7}$$

By realizing that $A^*AT^{3/2}e^{-q(\Phi_0+V_0)/k_BT}$ is the leakage current of the diode, $I_0$, and writing a new "effective" ideality constant $n_{Eff} = \frac{n}{1-nm}$, Eq. S7 is simplified to:

$$I_D = I_0\left[\exp\left(\frac{q}{n_{Eff}k_BT}V_D\right) - \exp\left(\frac{-q}{k_BT}mV_D\right)\right]. \tag{S8}$$

Eq. S8 is identical to the standard Shockley diode equation, but with a small correction for when the diode is reverse biased, $\exp\left(\frac{-q}{k_BT}mV_D\right)$. The ideality constant is replaced with an effective ideality constant whose value is tunable with the parameter $m$.

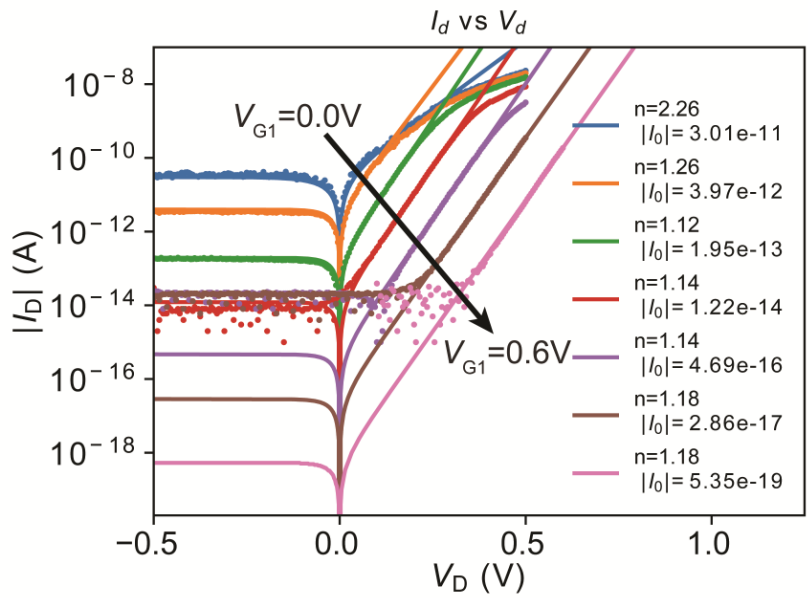

**Fig. S1. Extraction of reverse bias leakage currents**

Demonstration of the method used to extract the ideality factors and reverse bias leakage currents used in the main text. Due to the very flat characteristic of the reverse bias portion of our measurement, fitting the forward bias current is sufficient to accurately determine the diode leakage current.

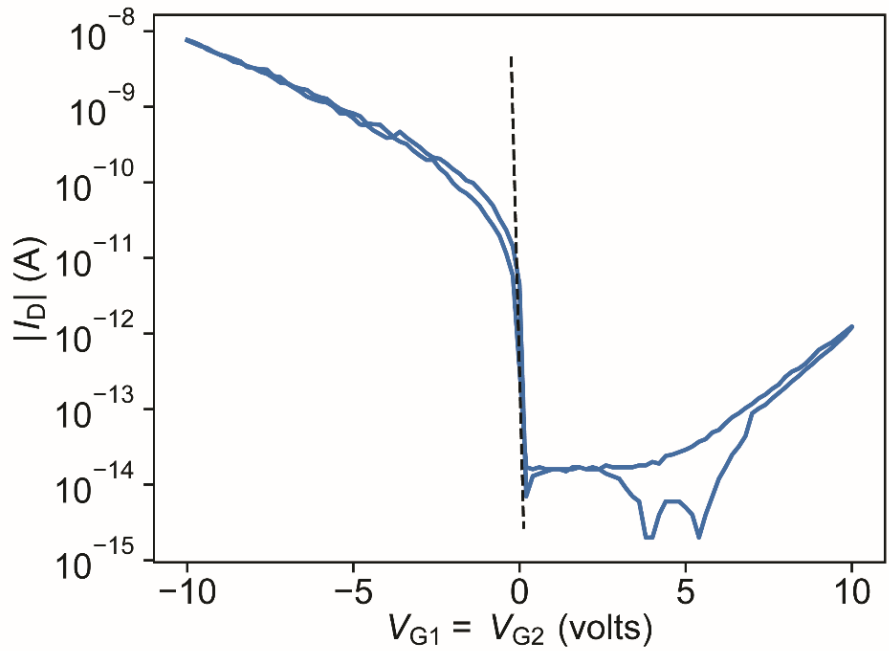

**Fig. S2. Device transfer curve**

Transfer curve measurement of the device described in the main text. The voltages of split-gates $V_{G1}$ and $V_{G2}$ are tied together, mimicking a global gate sweep to modulate the Fermi level of the entire channel. The drain voltage is fixed to -0.1V and the sample temperature is held at 300K throughout the measurement. The gate sweep direction goes from -10V to 10V and then back, demonstrating small measurement hysteresis. The dashed line is superimposed as a guide to the eye and has a slope representing the 60 mV/decade limit.

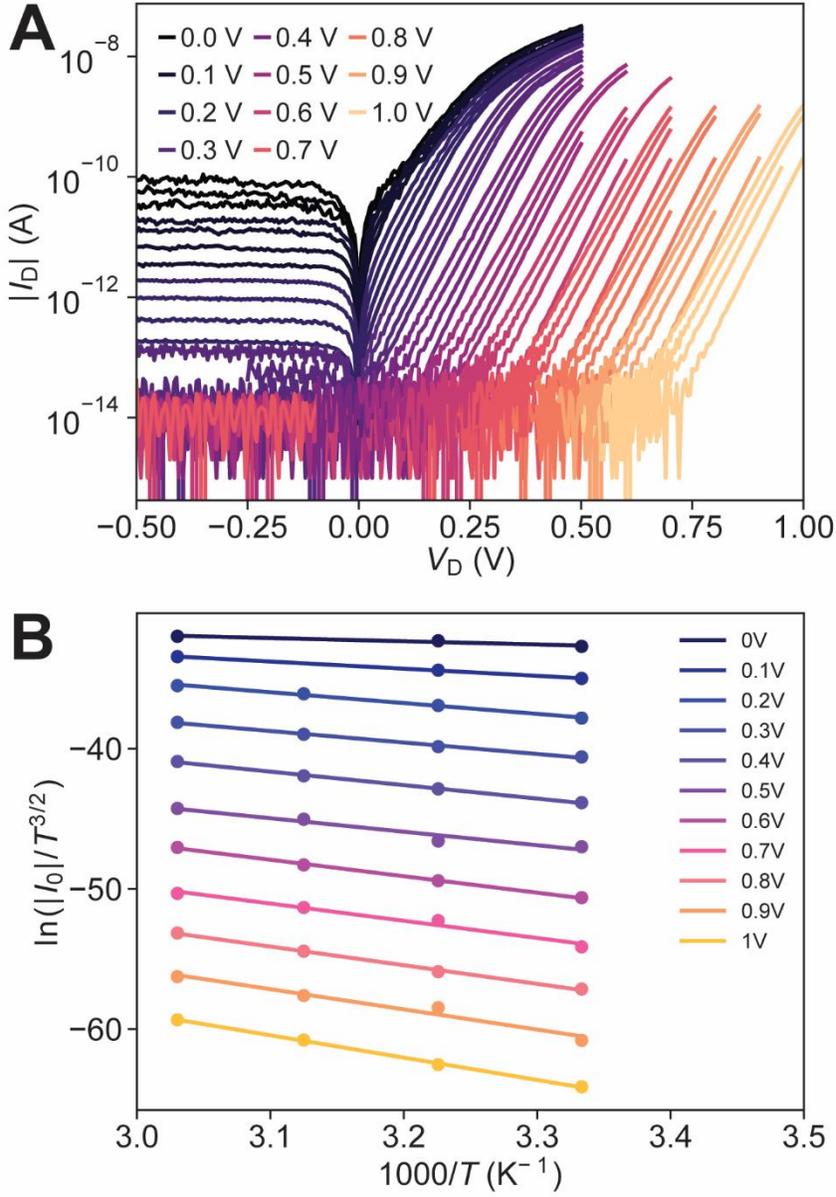

**Fig. S3. Temperature-dependent transport**

(**A**) Temperature-dependent measurements of a graphene-WSe$_2$ gated Schottky diode used in extraction of Schottky barrier heights. $V_{G2}$ is fixed to -10V while $V_{G1}$ is varied between 0.0 and 1.0V. Each measurement was performed at 300, 310, 320, and 330K, except for the $V_{G1} = 0.0$V and 0.1V data, which was performed at 300, 310, and 330K. Each $V_{G1}$ configuration is plotted as the same color, and as the temperature is increased, so does the current. (**B**) Arrhenius activation plot using Eq. S2 for the temperature-dependent data in Fig. S1a.

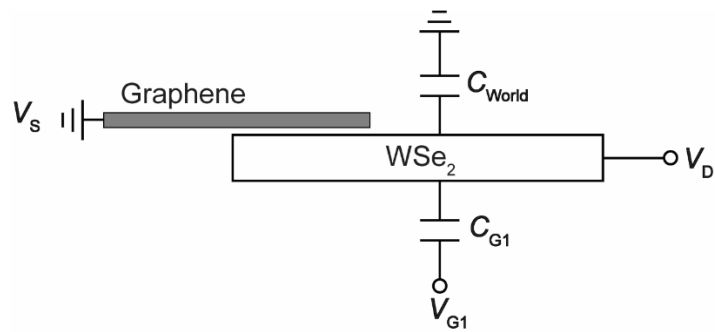

**Fig. S4. Voltage divider model for graphene-WSe₂ interface**

Measurement of graphene-WSe$_2$ GSD forward bias diode curve for $V_{G1}$ between 0.3 and 0.5V. Black arrows depict the different paths which may be followed during a simultaneous sweeping gate and drain voltage measurement. To observe $n_{Eff} < n$, the gate voltage must move in the opposite direction of the drain voltage. Conversely, to observe $n_{Eff} > n$, the gate voltage must be moved in the same direction as the drain voltage.

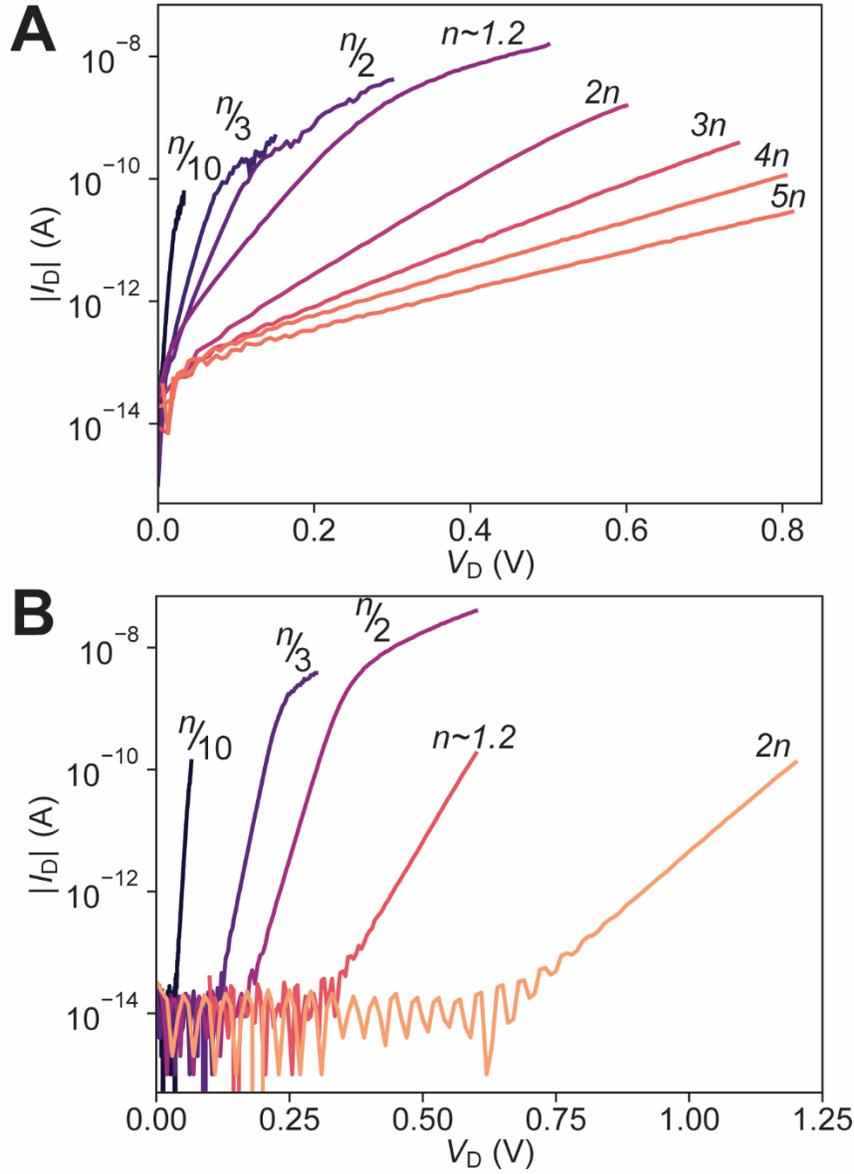

**Fig. S5. Super-ideal diode measurements for different starting $V_{G1}$ values**

Measurements of Gr-WSe$_2$ GSDs reconfigured as super-ideal diodes for different starting $V_{G1}$ values of **(A)** 0.2V and **(B)** 0.6V. As in Fig. 4 of the main text, to achieve $n_{Eff} < 1$, $V_{G1}$ is swept in the opposite direction as $V_D$. In order to avoid large voltages and stressing the device, values of $n_{Eff}$ beyond $2n$ were not measured for measurements starting at $V_{G1} = 0.6$V.

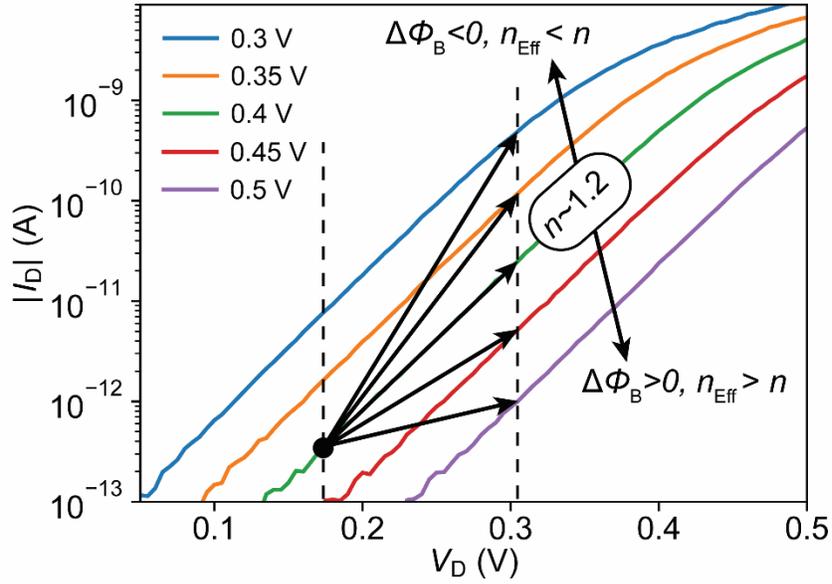

**Fig. S6. Paths for controlling $n_{\text{Eff}}$**

Measurement of graphene-WSe$_2$ GSD forward bias diode curve for $V_{G1}$ between 0.3 and 0.5V. Black arrows depict the different paths which may be followed during a simultaneous sweeping of gate and drain voltage measurement. To observe $n_{\text{Eff}} < n$, the gate voltage must move in the opposite direction of the drain voltage. Conversely, to observe $n_{\text{Eff}} > n$, the gate voltage must be moved in the same direction as the drain voltage.

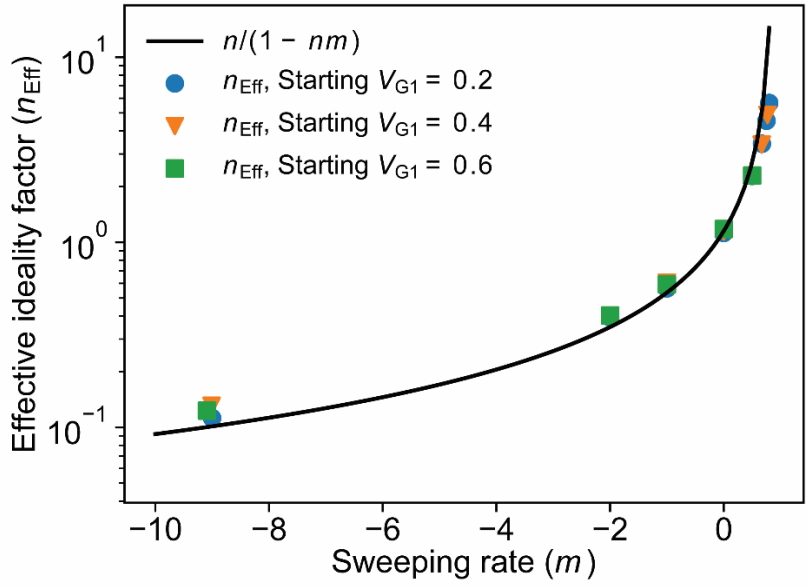

**Fig. S7. Comparison of measured $n_{\text{Eff}}$ with Eq. 1**

Here we demonstrate how $m$, the sweeping rate of the gate voltage relative to the drain voltage, can be used to control the effective ideality factor of the gated SB. The solid line is given by

$n_{\text{Eff}} = \dfrac{n}{1-nm}$ , where $n=1.15$. Scatter points are determined by fitting measured data from Fig. 4 and Fig. S5 to the diode equation.

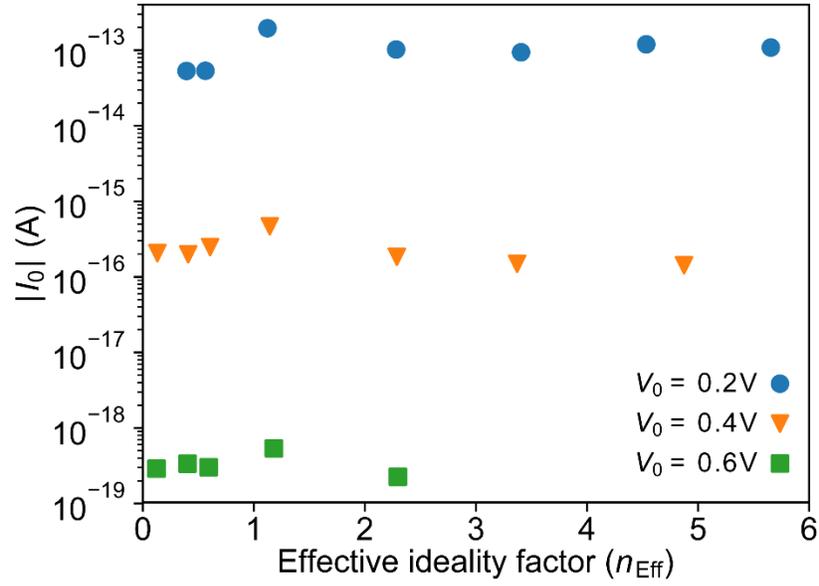

**Fig. S8. Comparison of measured $n_{\text{Eff}}$ with $I_0$**

$I_0$ and $n_{\text{Eff}}$ are both extracted by fitting data from Fig. 4 and Fig. S5. The reverse bias leakage current is primarily determined by $V_0$ and nearly is independent for different values of $n_{\text{Eff}}$, as is indicated by Eq. S8.

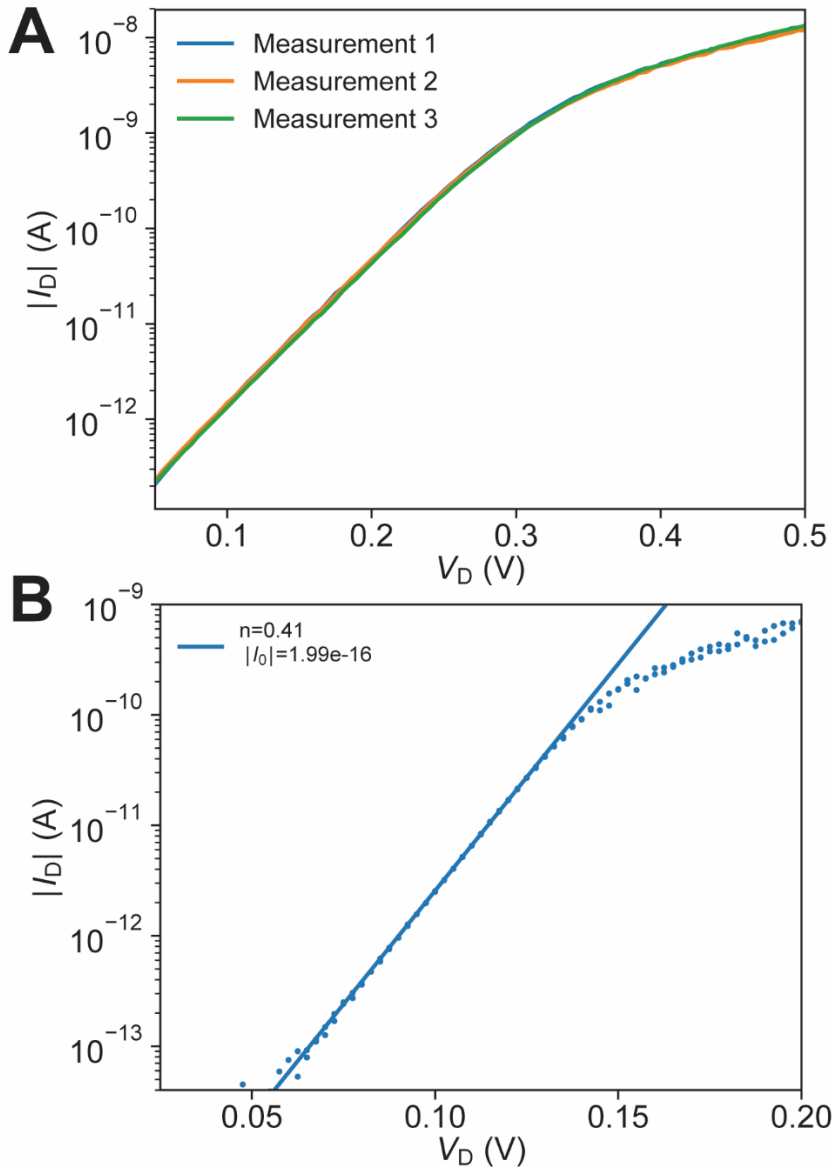

**Fig. S9. Measurement repeatability**

(**A**) $I_D V_D$ measurements with $V_{G1}$=0.25V, $V_{G2}$=-10V at a sample temperature of 300K repeated over twelve hours, with measurements in other gate configurations in between, demonstrating consistent results from measurement to measurement. (**B**) Forward and reverse sweep of super-ideal diode measurement for the $n_{\text{Eff}} = n/3$ configuration showed in Fig. 4 of the main text, demonstrating low measurement hysteresis during simultaneous sweeping of $V_{G1}$ and $V_D$. Scatter points are measured data and the solid line is a fit to the diode equation.